\documentclass{aa}
\usepackage{graphicx}

\begin{document}

\title{{\it XMM--Newton}\ observations of ultraluminous X--ray sources\\in nearby galaxies}

\author{L. Foschini\inst{1}, G. Di Cocco\inst{1}, L. C. Ho\inst{2}, L. Bassani\inst{1}, M. Cappi\inst{1}, M. Dadina\inst{1},
F. Gianotti\inst{1}, G. Malaguti\inst{1}, F. Panessa\inst{1}, E. Piconcelli\inst{1}, J.B. Stephen\inst{1}, M. Trifoglio\inst{1}}
\institute{Istituto di Astrofisica Spaziale e Fisica Cosmica (IASF--CNR) -- Sezione di Bologna\thanks{Formerly Istituto TeSRE -- CNR.},
Via Gobetti 101, I--40129, Bologna (Italy)
\and
The Observatories of the Carnegie Institution of Washington, 813 Santa Barbara Street, Pasadena, CA 91101 (USA)}

\offprints{L. Foschini, email: \texttt{foschini@tesre.bo.cnr.it}.}
\date{Received 6 May 2002; Accepted 24 June 2002}

\abstract{An \emph{XMM--Newton} study of ultraluminous X--ray sources (ULX) has been performed in a sample of 10
nearby Seyfert galaxies. Eighteen ULX have been found with positional uncertainty of about $4''$. The large
collecting area of \emph{XMM--Newton} makes the statistics sufficient to perform spectral fitting with simple models
in 8 cases. The main results of the present minisurvey strengthen the theory that the ULX could be accreting black
holes in hard or soft state. In some cases, the contribution of the ULX to the overall X--ray flux appears to be dominant
with respect to that of the active nucleus. In addition, 6 ULX present probable counterparts at other wavelengths (optical/infrared,
radio). A multiwavelength observing strategy is required to better assess the nature of these sources.
}

\titlerunning{Ultraluminous X--ray sources in nearby galaxies}
\authorrunning{L. Foschini et al.}

\maketitle

\section{Introduction}
The X--ray emission from Seyfert host galaxies comprises the contribution of a number of discrete
sources plus the hot interstellar plasma (Fabbiano 1989).
Most of the discrete sources appear to be close accreting binaries, with a compact companion.
\emph{Einstein} observations of the bulge of M31 revealed a population of about 100
low mass X--ray binaries (Fabbiano et al. 1987). Later, Supper et al. (1997) showed, by using
\emph{ROSAT} data, that the most luminous of these objects in M31 has $L_X=2\times 10^{38}$ erg s$^{-1}$, close to
the Eddington limit for a $1.4 M_{\sun}$ neutron star.

Recently, several sources with X--ray luminosities higher than the Eddington limit for a typical
neutron star have been detected in nearby galaxies (e.g., Read et al. 1997; Colbert \& Mushotzky 1999; Makishima et al. 2000; La Parola et al. 2001; Zezas et al. 2001). Fabbiano et al. (2001) found with
\emph{Chandra} 14 pointlike sources in the Antennae galaxies, with luminosities above $10^{39}$ erg s$^{-1}$
and up to $10^{40}$ erg s$^{-1}$.

These discoveries have raised difficulties in the interpretation of these sources. Even though it is statistically possible
to have \emph{some} individual cases of off--centre black holes with masses of the order of
$10^{3}-10^{4}$ $M_{\sun}$ (by assuming a typical Eddington ratio of $0.1-0.01$; cf. Nowak 1995),
it is very difficult to explain the high number of sources detected so far within this scenario.  Dynamical friction should have caused the objects to spiral
to the nucleus of the galaxy.
Several other hypotheses have been suggested about the nature of ULX: anisotropic emission from accreting black holes
(King et al. 2001), emission from jets in microblazars (K\"ording et al. 2002), emission from accreting Kerr
black holes (Makishima et al. 2000), and inhomogeneities in radiation--pressure dominated accretion disks (Begelman 2002).
However, the lack of sufficient information has not allowed us to distinguish between the different
models proposed. The search for optical counterparts has not yet yielded much data: to date only one ULX appears to have a
plausible counterpart (Roberts et al. 2001), and other ULX may be associated with planetary nebulae or
H~II regions (Pakull \& Mirioni 2002; Wang 2002).

Our team has been awarded about 250 ks of XMM--EPIC guaranteed time, and we started a distance--limited survey of
Seyfert galaxies. We selected $28$ objects in the northern hemisphere with $B_{\mathrm{T}}<12.5$ mag and $d<22$~Mpc (Di Cocco et al. 2000;
Cappi et al. 2002) from the Palomar survey of Ho et al. (1997a). The distances were estimated according to Ho et al. (1997a), and we adopt the same convention
in the present paper.

Here we present the results from a study of the discrete sources detected
in the galaxies, which are neither the nucleus nor background objects. To date we have obtained 13 objects
in our sample, but three observations were heavily corrupted by soft--proton flares and it
was not possible to extract any useful information. Here we present part of a study of the discrete source population in the
remaining 10 Seyfert galaxies (Table~\ref{tab:host}), specifically the catalog of ULX sources.
Some preliminary results have been presented in Foschini et al. (2002).

Previous detections of ULX have been largely confined to late--type 
galaxies (e.g., IC 342, M82, NGC 3628, and NGC 5204) or interacting systems 
undergoing a starburst phase (e.g., the Antennae). Although the objects studied
here technically have Seyfert nuclei, the level of nuclear activity is 
extremely low, and for the present purposes they can be considered 
``typical'' nearby galaxies.  The one selection effect to bear in mind is that 
most of the Palomar Seyferts tend to be relatively bulge--dominated disk
galaxies (see, e.g., Ho et al. 1997b), and so late--type galaxies are
underrepresented in our sample.

\begin{table*}[!ht]
\caption{Main characteristics of the observed host galaxies. Columns: (1) Name of the host galaxy from the New General Catalog;
(2) optical coordinates of the nucleus from Cotton et al. (2001); (3) Hubble type; (4) spectral
classification of the nucleus; (5) distance [Mpc]; (6) major axis of the $D_{25}$ ellipse [arcmin];
(7) Galactic absorption column density [$10^{20}$~cm$^{-2}$]; (8) date of \emph{XMM--Newton} observation [year--month--day];
(9) effective exposure time, i.e. cleaned from soft--proton flares [ks]. Data for Cols.~$3-6$ are taken from Ho et al. (1997a).}
\centering
\begin{tabular}{lcccccccc}
\hline
Galaxy & R.A., Dec. (J2000)  & Hubble Type & Sp. Class. & $d$ & $D_{25}$ & $N_{\mathrm{H}}$ & Date & Exp.\\
(1)    & (2)              & (3)         & (4)        & (5) & (6)      & (7)     & (8)        & (9)\\
\hline
NGC1058 & 02:43:30.2, +37:20:27.2     & SA(rs)c     & S2            & 9.1      & 3.02 &  6.65 & $2002-02-01$ & 6.0\\
NGC3185 & 10:17:38.7, +21:41:17.2     & SB(r)0/a    & S2        & 21.3     & 2.34 &  2.12 & $2001-05-07$ & 9.1\\
NGC3486 & 11:00:24.1, +28:58:31.6     & SAB(r)c     & S2            & 7.4      & 7.08 &  1.9  & $2001-05-09$ & 4.2\\
NGC3941 & 11:52:55.4, +36:59:10.5     & SB(s)0      & S2        & 18.9     & 3.47 &  1.9  & $2001-05-09$ & 5.0\\
NGC4138 & 12:09:29.9, +43:41:06.0     & SA(r)0+     & S1.9          & 17.0     & 2.57 &  1.36 & $2001-11-26$ & 10.0\\
NGC4168 & 12:12:17.3, +13:12:17.9     & E2          & S1.9      & 16.8     & 2.75 &  2.56 & $2001-12-04$ & 17.4\\
NGC4501 & 12:31:59.3, +14:25:13.4     & SA(rs)b     & S2            & 16.8     & 6.92 &  2.48 & $2001-12-04$ & 2.8\\
NGC4565 & 12:36:21.1, +25:59:13.5     & SA(s)b      & S1.9          & 9.7      & 15.85&  1.3  & $2001-07-01$ & 10.0\\
NGC4639 & 12:42:52.5, +13:15:24.1     & SAB(rs)bc   & S1.0          & 16.8     & 2.75 &  2.35 & $2001-12-16$ & 9.7\\
NGC4698 & 12:48:23.0, +08:29:14.8     & SA(s)ab     & S1.9          & 16.8     & 3.98 &  1.87 & $2001-12-16$ & 9.2\\
\hline
\end{tabular}
\label{tab:host}
\end{table*}

\begin{table*}[!ht]
\caption{ULX in the present catalog (significance greater than $4\sigma$). Columns: (1) Name of the host
galaxy; (2) ULX number; (3) coordinates of the ULX; (4) angular separation from the optical centre of the galaxy [arcsec];
(5) ULX name according to \emph{XMM-Newton} rules.}
\centering
\begin{tabular}{lcccc}
\hline
Host Galaxy & Object & RA, Dec (J2000.0)  & Separation & {\it XMM} ID\\
(1)         & (2)    & (3)                & (4)      & (5)\\
\hline
NGC1058 & ULX1    & 02:43:23.5, +37:20:38           & 77           & XMMU J024323.5+372038 \\
{}      & ULX2    & 02:43:28.3, +37:20:23           & 19           & XMMU J024328.3+372023 \\
NGC3185 & ULX1    & 10:17:37.4, +21:41:44           & 30           & XMMU J101737.4+214144 \\
NGC3486 & ULX1    & 11:00:22.4, +28:58:18           & 23           & XMMU J110022.4+285818 \\
NGC3941 & ULX1    & 11:52:58.3, +36:59:00           & 38           & XMMU J115258.3+365900 \\
NGC4168 & ULX1    & 12:12:14.5, +13:12:48           & 45           & XMMU J121214.5+131248 \\
NGC4501 & ULX1    & 12:32:00.1, +14:22:28           & 166          & XMMU J123200.1+142228 \\
{}      & ULX2    & 12:32:00.8, +14:24:42           & 40           & XMMU J123200.8+142442 \\
NGC4565 & ULX1    & 12:36:05.2, +26:02:34           & 289          & XMMU J123605.2+260234 \\
{}      & ULX2    & 12:36:14.8, +26:00:53           & 127          & XMMU J123614.8+260053 \\
{}      & ULX3    & 12:36:17.3, +25:59:51           & 59           & XMMU J123617.3+255951 \\
{}      & ULX4    & 12:36:17.4, +25:58:54           & 51           & XMMU J123617.4+255854 \\
{}      & ULX5    & 12:36:18.8, +26:00:34           & 83           & XMMU J123618.8+260034 \\
{}      & ULX6    & 12:36:27.8, +25:57:34           & 139          & XMMU J123627.8+255734 \\
{}      & ULX7    & 12:36:30.6, +25:56:50           & 197          & XMMU J123630.6+255650 \\
NGC4639 & ULX1    & 12:42:48.3, +13:15:41           & 61           & XMMU J124248.3+131541 \\
{}      & ULX2    & 12:42:51.4, +13:14:39           & 50           & XMMU J124251.4+131439 \\
NGC4698 & ULX1    & 12:48:25.9, +08:30:20           & 73           & XMMU J124825.9+083020 \\
\hline
\end{tabular}
\label{tab:srccoord}
\end{table*}

\begin{table*}[!ht]
\caption{X--ray data on the ULX of the present catalog.
Columns: (1) Host galaxy; (2) ULX number; (3) count rate in counts per second, as calculated with \emph{eboxdetect}; (4)
likelihood of the detection (cf. Ehle et al. 2001); (5) absorbing column density [$10^{21}$ cm$^{-2}$]; (6) photon index
of the power law;
(7) X--ray luminosity in the $0.5-10$ keV energy band [$10^{38}$ erg s$^{-1}$], calculated from the count rates of Col.~3
and converted with the factor of $3\times 10^{11}$~cnt$\cdot$cm$^2$/erg, which in turn is derived by using the
power-law model with $\Gamma=2.0$ and an average Galactic $N_{\mathrm{H}}=3\times 10^{20}$ cm$^{-2}$ (cf. Ehle et al. 2001). When the parameters $N_H$ and
$\Gamma$ are present, it means that the statistics are sufficient to perform a spectral fitting and uncertainties in the parameters
estimate are at the $90\%$ confidence limits. For $N_{\mathrm{H}}=N_{\mathrm{H, gal}}$, it means that no additional absorption is found to be significant.
The luminosities were calculated using distances from Ho et al. (1997a), who considered an infall velocity of 300 km/s for the Local Group,
$H_0=75$~km$\cdot$s$^{-1}$Mpc$^{-1}$, and the distance of the Virgo Cluster of 16.8~Mpc (cf. Table~1, Col.~5).}
\centering
\begin{tabular}{lcccccc}
\hline
Host Galaxy & Object & Count Rate $[10^{-3}]$  & Likelihood & $N_{\mathrm{H}}$  & $\Gamma$ & $L_{0.5-10 keV}$\\
(1)         & (2)    & (3)                              & (4)        & (5)      & (6)      & (7)\\
\hline
NGC1058 & ULX1    & $13\pm 2$                           & 79         & $N_{\mathrm{H, gal}}$ & $1.1\pm 0.3$  & $11$\\
{}      & ULX2    & $6\pm 1$                            & 22         & --       & --       & $2.3$\\
NGC3185 & ULX1    & $5\pm 1$                            & 27         & --       & --       & $13$\\
NGC3486 & ULX1    & $20\pm 4$                           & ($^{\mathrm{*}}$) & $N_{\mathrm{H, gal}}$ &  $2.2\pm 0.5$ & $5.0$\\
NGC3941 & ULX1    & $43\pm 4$                           & 320        &  $N_{\mathrm{H, gal}}$ & $1.9\pm 0.2$ & $74$\\
NGC4168 & ULX1    & $4.0\pm 0.7$                        & 36         & --       & --       & $6.0$\\
NGC4501 & ULX1    & $11\pm 3$                           & 23         & --       & --       & $17$\\
{}      & ULX2    & $29\pm 4$                           & 93         & $N_{\mathrm{H, gal}}$ & $2.3\pm 0.4$  & $37$\\
NGC4565 & ULX1    & $7\pm 1$                            & 44         & --       & --       & $3.4$\\
{}      & ULX2    & $10\pm 1$                           & 90         & $6\pm 5$ & $1.7\pm 0.6$ & $16$\\
{}      & ULX3    & $3.9\pm 0.9$                        & 21         & --       & --       & $2.0$\\
{}      & ULX4    & $51\pm 3$                           & 821        & $N_{\mathrm{H, gal}}$ & $1.9\pm 0.1$  & $25$\\
{}      & ULX5    & $7\pm 1$                            & 57         & --       & --       & $3.4$\\
{}      & ULX6    & $13\pm 2$                           & 121        & $N_{\mathrm{H, gal}}$ & $1.5\pm 0.3$  & $9.0$\\
{}      & ULX7    & $4\pm 1$                            & 23         & --       & --       & $2.0$\\
NGC4639 & ULX1    & $5\pm 1$                            & 29         & --       & --       & $8.0$\\
{}      & ULX2    & $3\pm 1$                            & ($^{\mathrm{*}}$)  & -- & --     & $5.0$\\
NGC4698 & ULX1    & $16\pm1$                            & 168        & $N_{\mathrm{H, gal}}$ & $2.0\pm 0.2$     & $30$\\
\hline
\end{tabular}
\begin{list}{}{}
\item[$^{\mathrm{*}}$] Data from manual analysis.
\end{list}
\label{tab:srcdata}
\end{table*}

\section{XMM--Newton data analysis}
The present ULX catalog is based on data from the European Photon Imaging Camera (EPIC) on board the \emph{XMM--Newton}
satellite. EPIC comprises three instruments: the PN--CCD camera (Str\"uder et al. 2001) and
two MOS--CCD detectors (Turner et al. 2001).

To define a ULX source as an off--nuclear galactic object, we use three selection criteria:

\begin{itemize}
\item the object luminosity, in the energy band $0.5-10$~keV, must be greater than $2\times 10^{38}$~erg s$^{-1}$;

\item the object has to be located inside the $D_{25}$ ellipse (i.e., the dimension equal to 25 mag/arcsec$^2$) of the
host galaxy;

\item the object must be sufficiently far from the optical centre, to avoid confusion with the galaxy's active nucleus.

\end{itemize}

Considering that the absolute location accuracy for \emph{XMM--Newton} is about $4''$ (Jansen et al. 2001),
and that the uncertainty in the optical position of the host galaxy centre, from the Digitized Sky Survey
(DSS, see Cotton et al. 1999) is less than $2.7''$ , we searched for off--nuclear sources at least $10''$ away from
the optical centre of the host galaxy. The characteristics of the sources in the present
catalog are summarized in Table~\ref{tab:srccoord}.

For the processing, screening, and analysis of the data we used the standard tools
of XMM--SAS software v. 5.2 and HEAsoft Xspec (11.0.1). The images are prepared
with DS9 v. 2.1, together with ZHTools v. 2.0.2.

Soft--proton flares affected the observations randomly, and in some cases it is necessary to
filter the available data. Time intervals contaminated by flares have been excluded by extracting
the background lightcurve in the $10-13$~keV energy band. Periods with count rates higher
than $0.2$~s$^{-1}$ have been removed.

\begin{figure*}[ht]
  \begin{center}
    \includegraphics[width=8cm]{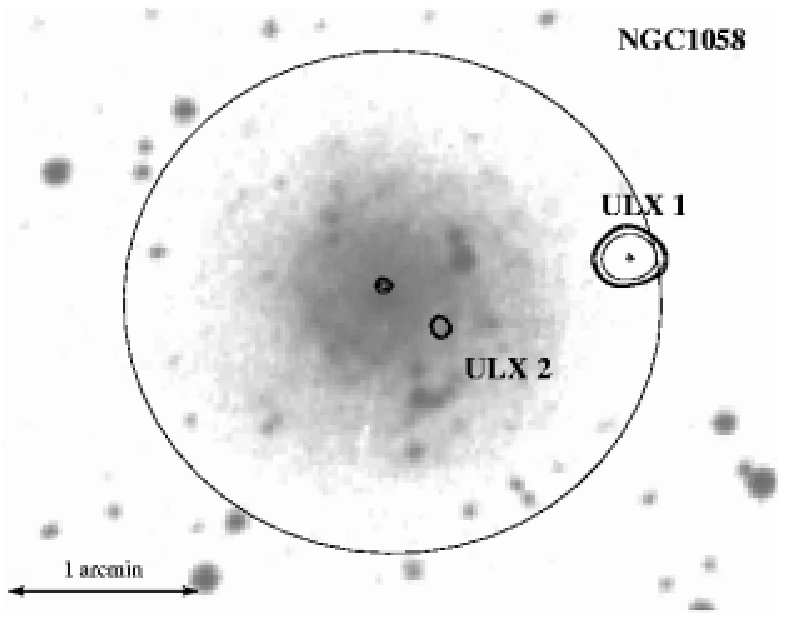}
    \hspace{12pt}
    \includegraphics[width=8cm]{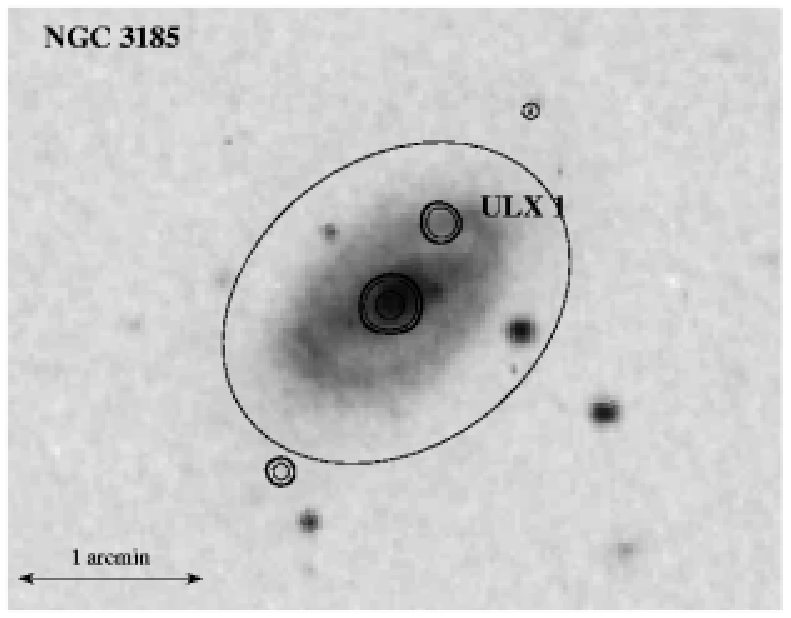}
  \end{center}

  \begin{center}
    \includegraphics[width=8cm]{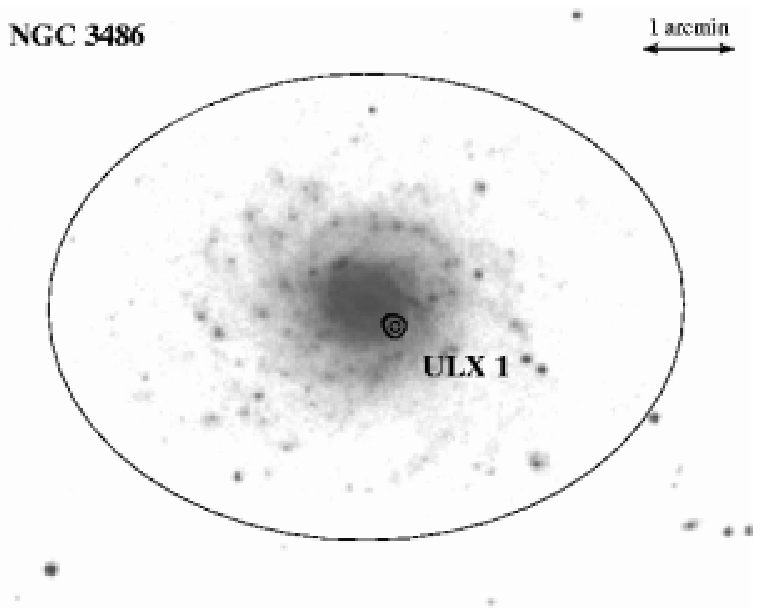}
    \hspace{12pt}
    \includegraphics[width=8cm]{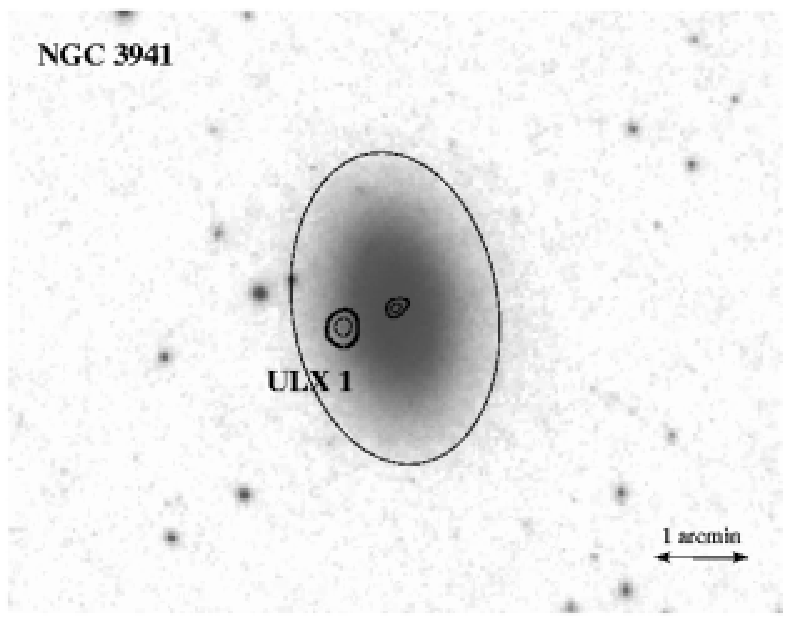}
  \end{center}
\caption{Images from Digitized Sky Survey (DSS) with \emph{XMM--Newton} data superimposed. X--ray contours are
calculated from MOS images in the $0.5-10$~keV energy band. North is up and East to the left.
The $D_{25}$ ellipse is also shown for comparison.}
\label{fig1}
\end{figure*}

To perform the detection, the EPIC--PN was selected, because of its larger effective area
with respect to the MOS cameras, which allows more accuracy in the detection of faint X-ray sources.
The detections were performed using the sliding box cell detection algorithm (\emph{eboxdetect} of XMM--SAS).
It uses a box, with dimension $5\times 5$ pixel as the detection cell,
and $L=-\ln(P)=10$ as the minimum detection likelihood value, which in turn corresponds to
a probability of Poissonian random fluctuations of the counts in the detection cell
of $P=4.5 \times 10^{-5}$ (roughly $4\sigma$).

After the automatic procedure, each source inside the $D_{25}$ ellipse was carefully checked to exclude
false or doubtful sources. Specifically, we noted that \emph{eboxdetect} fails in some cases.
Indeed, the software accumulates the source counts in a $5\times 5$ pixel box, while the background
counts are accumulated in the region of $(9\times 9) - (5\times 5)$ pixels. This algorithm gives good results
with uniform regions (both diffuse or with background only), but fails in border regions. By comparing visually
the detections from EPIC--PN with data from the MOS cameras, it is possible to identify and exclude possible artifacts.
It is worth noting that in the case of fake detection, \emph{eboxdetect} gives an unusually poor point source
location accuracy (PSLA) of $3-4''$, to be compared with a PSLA of less than $0.1''$
for real and normal detections.  (Note that the global positional uncertainty is given by the sum of the
satellite pointing uncertainty of $4''$ and the PSLA.)

After the detection run, we then extracted from the available list only those sources with X--ray luminosities higher
than $2\times 10^{38}$~erg s$^{-1}$ in the energy band $0.5-10$~keV (see Table \ref{tab:srccoord}).

The count rate calculated by \emph{eboxdetect} has been converted into flux using a conversion factor of
$3\times 10^{11}$~cnt$\cdot$cm$^2$/erg. This has been calculated from the graphics available in the
\emph{XMM--Newton User's Handbook} (Ehle et al. 2001) and by assuming a power law with photon index
2 and an absorbing column density $N_{\mathrm{H}}=3.0\times 10^{20}$~cm$^{-2}$. This choice of the model parameters
is consistent with results obtained from the spectral analysis of the brightest sources (see the next section).

We do not apply the correction for vignetting, because all the sources for which it is possible to extract the spectrum
are close to the centre of the field of view (less than $2'$) and have most of their statistics
below $5$~keV (cf. Lumb 2002).

\begin{figure*}[ht]
  \begin{center}
    \includegraphics[width=8cm]{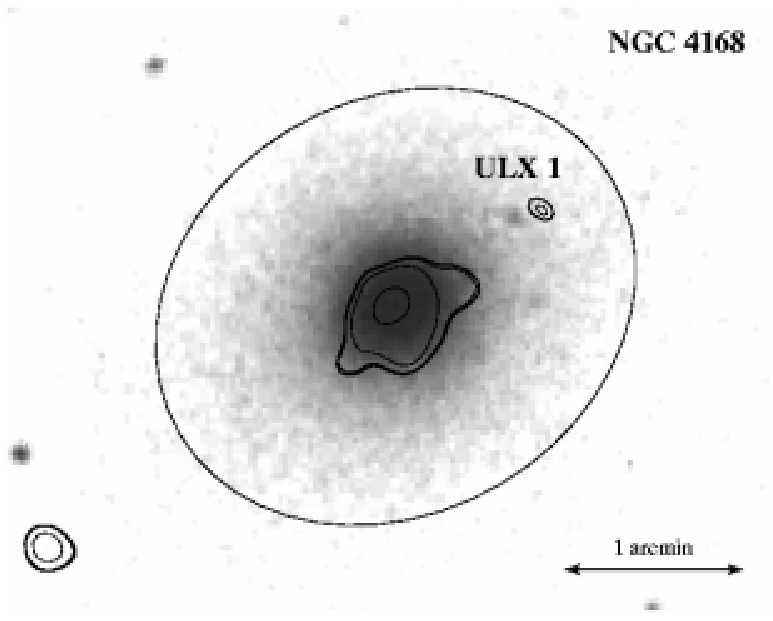}
    \hspace{12pt}
    \includegraphics[width=8cm]{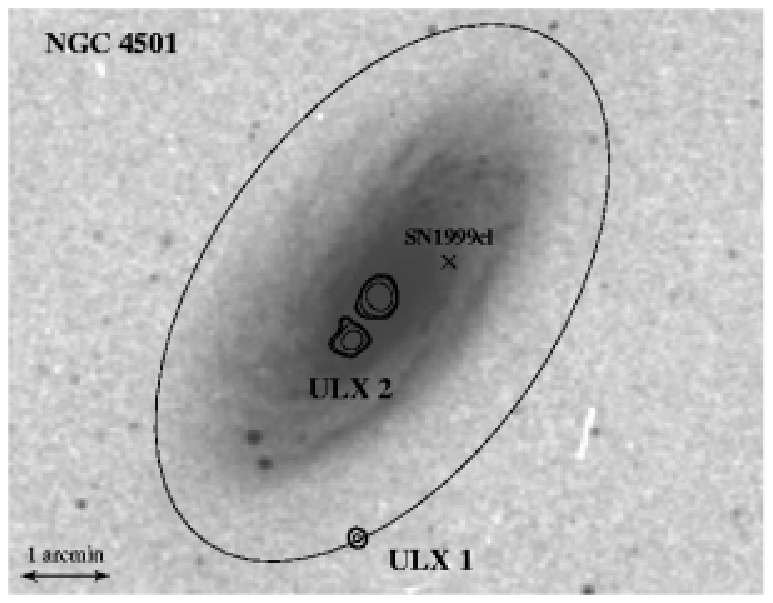}
  \end{center}

  \begin{center}
    \includegraphics[width=8cm]{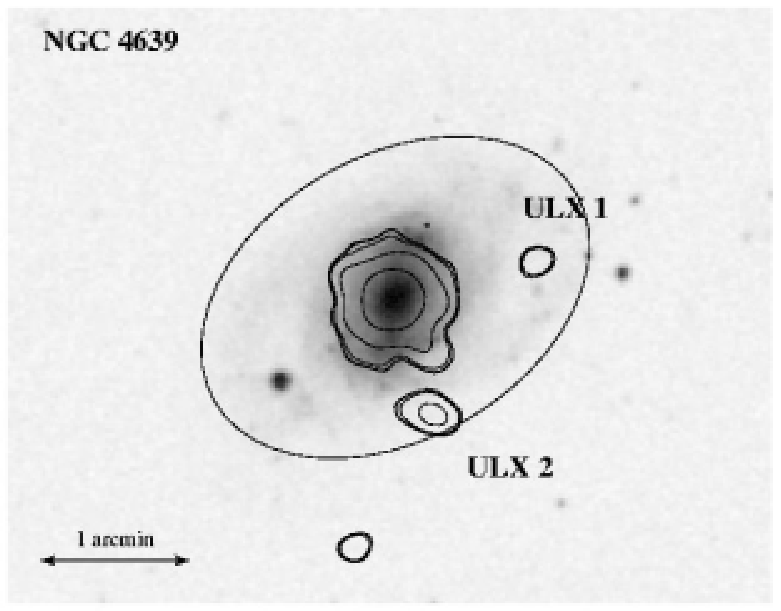}
    \hspace{12pt}
    \includegraphics[width=8cm]{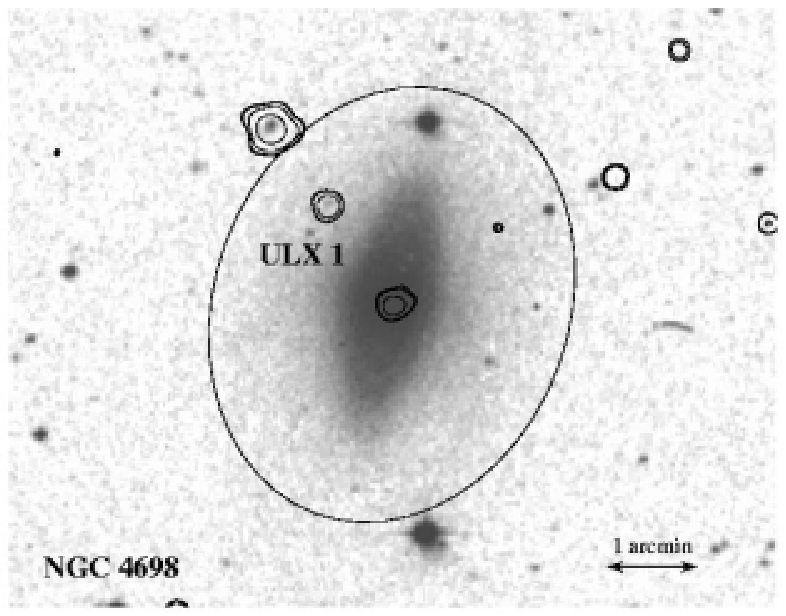}
  \end{center}

\caption{Images from Digitized Sky Survey (DSS) with \emph{XMM--Newton} data superimposed. X--ray contours are calculated from MOS images
in the $0.5-10$~keV energy band. North is up and East to the left. The $D_{25}$ ellipse is also shown for comparison.}
\label{fig2}
\end{figure*}

\begin{figure*}[ht]
  \begin{center}
    \includegraphics[width=10cm]{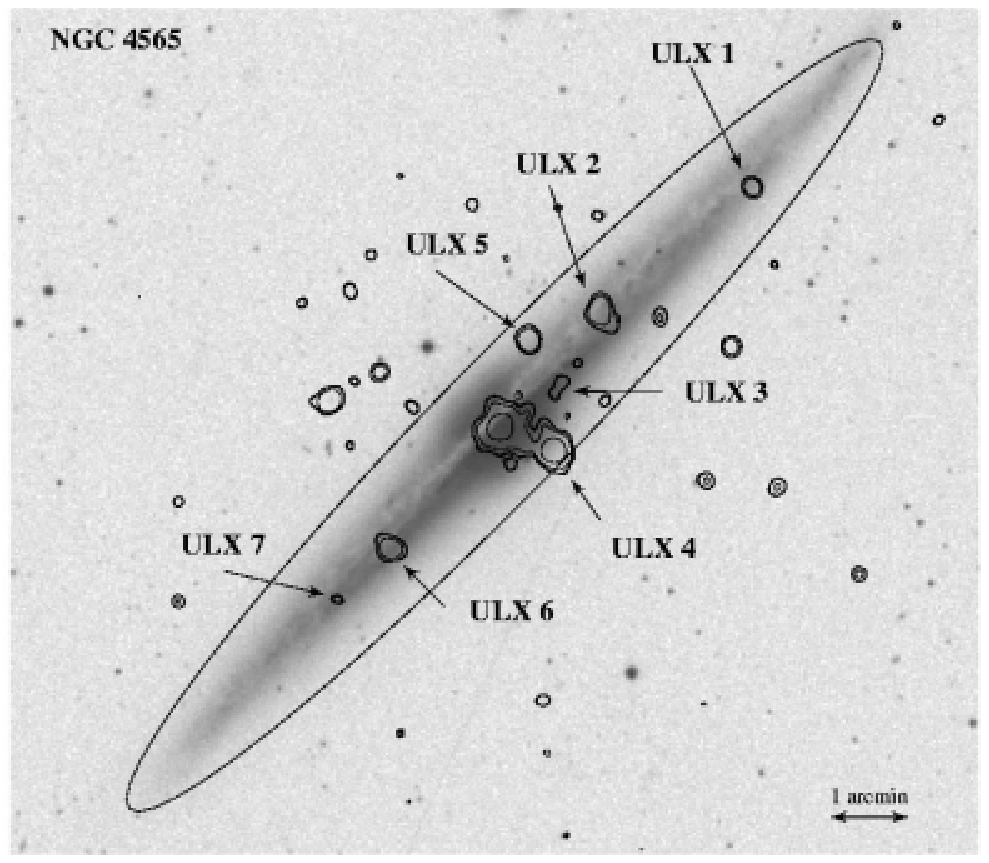}
  \end{center}

\caption{Image from Digitized Sky Survey (DSS) with \emph{XMM--Newton} data superimposed. X--ray contours are calculated from MOS images
in the $0.5-10$~keV energy band. North is up and East to the left. The $D_{25}$ ellipse is also shown for comparison.}
\label{fig3}
\end{figure*}

\section{Analysis of the X--ray sources}
Figs~$1-3$ show DSS images superimposed on the smoothed X--ray contour
plots (data from EPIC--MOS2 camera). The $D_{25}$ ellipse is also
shown for comparison. Here and in the following, if not explicitly indicated, the data on
the host galaxies are obtained from the catalog of Ho et al. (1997a).

The X--ray properties of the ULX are given in Table \ref{tab:srcdata}.
For eight of these sources the statistics are good enough (at least $80$ counts) to perform a spectral
fitting with simple models. To extract the spectrum we selected source regions
with radius $20'' - 30''$, depending on the presence or absence 
of nearby sources. In all cases, the flux was corrected according to the 
encircled energy fraction (Ghizzardi 2001).

Data from MOS1, MOS2, and PN were fitted simultaneously. We used the following models: power law (PL), black body
(BB), thermal bremstrahlung (BR), unsaturated Comptonization (CST) by
Sunyaev \& Titarchuk (1980), and the multicolor black body accretion disk (MCD) by Mitsuda et al. (1984). The latter two 
correspond to the \emph{compST} and \emph{diskbb} models in \emph{xspec}. The unsaturated Comptonization model has two free
parameters, the temperature and the optical depth. The multicolor disk has one free parameter, i.e. the temperature
at the inner disk radius.

In Table~\ref{tab:srcdata} are summarized the results for the power-law model, while in the following we discuss other cases.
We consider acceptable only the fits with reduced  $\chi^2$ less than 2.

\subsection{NGC1058}
Two ULX have been detected and both of them appear to have no evident optical counterpart in the DSS. ULX1 has sufficient counts to be
fitted. The best model is the power law, even though with an unusual low $\Gamma = 1.1\pm 0.3$ ($\chi^2=41.1$, $\nu=33$). The
$0.5-10$~keV flux is calculated as $8.5\times 10^{-14}$~erg cm$^{-2}$ s$^{-1}$, which corresponds to a luminosity
of $1.1\times 10^{39}$~erg s$^{-1}$. Other models give worse fits; for 
example, the black body model with $kT=0.7\pm 0.2$~keV gives $\chi^2=53.0$ for 
$\nu=33$, and we get an upper limit for the temperature of the inner disk in the MCD model of $kT<4$~keV ($\chi^2=43.8$, $\nu=33$).

\subsection{NGC3185}
This source is the farthest in our sample, and inside its small angular size ($D_{25}=2.34'$),
we find one clear ULX, for which the low statistics do not allow a spectral fit. 
No clear optical counterpart
is visible in the DSS.

\subsection{NGC3486}
One ULX has been detected at $23''$ from the optical centre. The statistics are sufficient
to perform a spectral fitting. The best-fit model is obtained with a simple power law model
with a photon index of $2.2\pm 0.5$ ($\chi^2=7.7$, $\nu=14$).
The $0.5-10$~keV flux is $8.3\times 10^{-14}$~erg cm$^{-2}$ s$^{-1}$, which corresponds to a luminosity
of $5\times 10^{38}$~erg s$^{-1}$.
The fits are worse, but still acceptable, with the BB ($\chi^2=13.3$, $\nu=14$) and MCD ($\chi^2=11.9$, $\nu=14$) models.
The former gives a temperature of $kT=0.26\pm 0.05$~keV, while the latter gives a temperature of the inner disk
of $kT=0.4\pm 0.2$~keV.

After the analysis of 41.3 ks of {\it ASCA}/SIS0 data, Pappa et al. (2001) found, for this galaxy, an observed flux
of $5\times 10^{-14}$~erg cm$^{-2}$ s$^{-1}$ (in the energy band $0.8-10$~keV), using a power
law with $\Gamma=1.9$ absorbed by an additional column density of $3.2\times 10^{21}$~cm$^{-2}$.
They suggested that NGC3486 may be an obscured Seyfert 2 galaxy.
However, from \emph{XMM-Newton} data, in a circle of $10''$ centered in the optical centre of NGC3486,
the count rate is $(2.1\pm 0.7)\times 10^{-3}$~s$^{-1}$, which corresponds to a flux of 
$(9\pm 3)\times 10^{-15}$~erg cm$^{-2}$ (given the conversion factor used in this paper).
The luminosity is $(6\pm 2)\times 10^{37}$~erg s$^{-1}$ in the $0.5-10$~keV energy band. These data are compatible
with those of \emph{Chandra} (Ho et al. 2001) and the upper limit of {\it ROSAT}/HRI
($6.1\times 10^{-14}$~erg cm$^{-2}$ s$^{-1}$ in the $0.1-2.4$~keV band; Halderson et al. 2001).

Since the {\it ASCA} flux was calculated by extracting photons in wider regions ($1.5'$), because of the low angular
resolution, this suggest the source observed by {\it ASCA} was not the nucleus, but the ULX, at only $23''$ ($\sim$825~pc)
from the optical centre of NGC3486.

\subsection{NGC3941}
Only one ULX has been detected in this galaxy at $38''$ from the centre. The spectrum is well fitted
with a PL model with $\Gamma=1.9\pm 0.2$ ($\chi^2=17.8$, $\nu=19$) or a BR model with $kT=4\pm 2$~keV ($\chi^2=24.4$, $\nu=19$).
In the first case, the flux is $1.7\times 10^{-13}$~erg cm$^{-2}$ s$^{-1}$ and the luminosity
is $7.4\times 10^{39}$~erg s$^{-1}$.

\subsection{NGC4138}
No ULX was detected inside the $D_{25}$ ellipse (figure not shown). The flux limit is about
$10^{-14}$~erg cm$^{-2}$ s$^{-1}$ for this observation.

\subsection{NGC4168}
The \emph{XMM--Newton} observation shows one clear ULX in this galaxy, which appears to have a possible correlation
with a point source in the 2MASS survey. No spectral fitting was possible.

\subsection{NGC4501}
We clearly detected two ULX in this galaxy, one of them just on the border of the $D_{25}$ ellipse.
ULX2 has sufficient counts for a spectral fitting. The best fit ($\chi^2=26.3$, $\nu=26$) is a power law with
$\Gamma = 2.3\pm 0.4$, which corresponds to a flux of $1.0\times 10^{-13}$~erg cm$^{-2}$ s$^{-1}$ and a luminosity of 
$3.7\times 10^{39}$~erg s$^{-1}$. Other acceptable models are BB ($\chi^2=40.9$, $\nu=26$) with
$kT=0.28\pm 0.09$ keV and MCD ($\chi^2=35.6$, $\nu=26$) with $kT=0.5\pm 0.3$ keV. For BR we obtain only an upper limit for
$kT<4$ keV ($\chi^2=30.2$, $\nu=26$).

\subsection{NGC4565}
We find seven sources inside the $D_{25}$ ellipse, but since this galaxy is edge--on,
there could be additional sources projected along the minor axis
above the plane of the disk.  Examination of Fig.~3 suggests that this
indeed might be the case, although interestingly none of the sources
seem to coincide with known globular clusters (Kissler-Patig et al. 1999).

For three ULX (2, 4, and 6) it is possible to perform spectral fitting. Specifically, ULX4 has the highest
counts in all the present catalog. The best fit is still the power law with photon index $1.9\pm 0.1$
($\chi^2=43.7$, $\nu=46$). The flux and the corresponding luminosity are
$2.2\times 10^{-13}$~erg cm$^{-2}$ s$^{-1}$ and $2.5\times 10^{39}$~erg s$^{-1}$, respectively.

It is worth noting that, with the exception of the BB model, all of the 
other models give an acceptable spectral fit for ULX4.
In addition, it is also the only source that is fitted well with the unsaturated Comptonization model.
Specifically, the BR model is fitted with $kT=3.4\pm 0.8$ keV ($\chi^2=48.4$, $\nu=46$), the MCD with $kT=0.8\pm 0.1$ keV
($\chi^2=83.1$, $\nu=46$), and the CST has $kT=1.8\pm 0.6$ keV with optical depth
$\tau_e = 16\pm 7$ ($\chi^2=42.8$, $\nu=45$).

ULX2 is best fitted with the black body model at $kT=0.9\pm 0.1$~keV ($\chi^2=17.7$, $\nu=13$).
The flux and luminosity for this model are $7.5\times 10^{-14}$~erg cm$^{-2}$ s$^{-1}$ and
$8\times 10^{38}$~erg s$^{-1}$, respectively. Other reasonable fits are obtained with PL
(see Table \ref{tab:srcdata}) and with MCD, with the temperature of the inner
disk of $2\pm 1$ keV ($\chi^2=20.8$, $\nu=13$).

ULX6 also is best fitted with BB, but with $kT=0.52\pm 0.07$ keV ($\chi^2=13.5$, $\nu=10$).
In this case the flux is $3.9\times 10^{-14}$~erg cm$^{-2}$ s$^{-1}$ and the luminosity is 
$4\times 10^{38}$~erg s$^{-1}$. Yet other statistically acceptable models are the PL (see Table \ref{tab:srcdata})
and the MCD, with temperature of the inner disk of $1.0\pm 0.3$ keV ($\chi^2=15.1$, $\nu=10$).

Four of the ULX sources have already been detected by {\it ROSAT} (Vogler et al. 1996),
namely $RXJ1236.2+2600$ (ULX2), $RXJ1236.2+2558$ (ULX4), $RXJ1236.3+2600$ (ULX5), $RXJ1236.4+2557$ (ULX6).
For ULX4, {\it ROSAT} observations suggest an additional intrinsic absorption, with values from $3.4$ to
$3.6\times 10^{20}$~cm$^{-2}$, in addition to the Galactic column density ($1.3\times 10^{20}$~cm$^{-2}$).
Instead, the best fit from {\it ASCA} data gives an upper limit of $2\times 10^{20}$~cm$^{-2}$ (Mizuno et al. 1999).
In our sample, only ULX2 indicates a possible extra absorption in addition to the Galactic $N_H$, with 95\% significance
(cf. Table \ref{tab:srcdata}). Additional absorption in the fitting of ULX4 and ULX6 has low significance (68\% and 82\%,
respectively). The values of $N_{\mathrm{H}}$ are $< 1\times 10^{21}$~cm$^{-2}$ and $2\times 10^{21}$~cm$^{-2}$, respectively.

It is useful to note that Mizuno et al. (1999) suggest, on the basis of {\it ASCA} data, that this galaxy has no X-ray nucleus and that the twin bright
sources in the middle of the galaxy could be two ULX. \emph{XMM-Newton} has sufficient angular resolution
to separate the two sources, one of them being ULX4; this is consistent with 
the {\it ROSAT} results. We identify the other source with the active
nucleus.

\subsection{NGC4639}
We find two ULX, but one of them (ULX2) was detected with the MOS, but not
with the PN, because its position fell into a gap between the PN chips. In both cases,
the statistics were not sufficient to perform spectral fitting.

\subsection{NGC4698}
We find one ULX with sufficient statistics to perform spectral fitting.
The best fit is a power law with $\Gamma = 2.0\pm 0.2$ ($\chi^2=18.3$, $\nu=24$)
that gives a flux of $8.6\times 10^{-14}$~erg cm$^{-2}$ s$^{-1}$ and a
luminosity of $3\times 10^{39}$~erg s$^{-1}$.

Other statistically acceptable models are BB with $kT=0.35\pm 0.05$~keV ($\chi^2=34.5$, $\nu=24$),
BR with $kT=2\pm 1$~keV ($\chi^2=19.0$, $\nu=24$), and
MCD with $kT=0.6\pm 0.1$ keV ($\chi^2=23.7$, $\nu=24$).

\begin{figure*}[!ht]
  \begin{center}
    \includegraphics[angle=270, width=7.5cm]{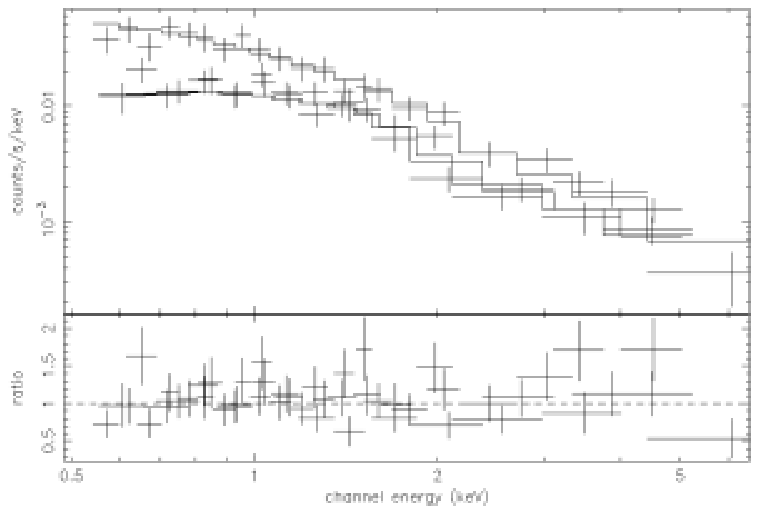}
    \hspace{12pt}
    \includegraphics[angle=270, width=7.5cm]{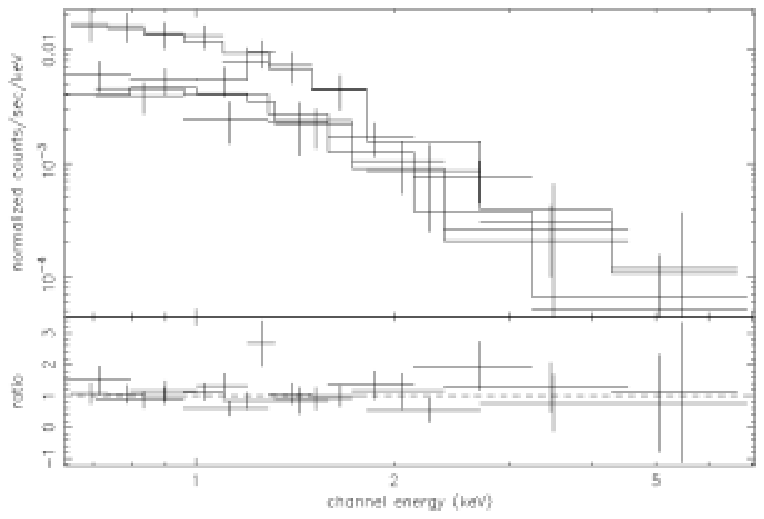}
  \end{center}
\caption{Examples of spectra. Combined spectrum (MOS1, MOS2, PN) of the source
NGC4565--ULX4 ({\it left}) and the source NGC4698--ULX1 ({\it right}). Both were fitted with a single power-law model.
Refer to Table \ref{tab:srcdata} for more details. The lower window of the two panels shows 
the ratio between the data and the model.}
\label{fig4}
\end{figure*}

\section{Counterparts at other wavelengths}
We performed a search for possible counterparts at other wavelengths using available online
databases (NED and Simbad). Results from a more extensive search are beyond the scope of the
present paper and will be presented in a future work.

We find a point source in the online 2MASS catalog at $1.72''$ from NGC4168--ULX1. Magnitudes are
$16.8\pm 0.2$, $16.1\pm 0.1$, and $15.3\pm 0.1$ in the $J$, $H$, and $K$ bands,  respectively.
ULX1 in NGC4168 also has a possible counterpart, $2.7''$ away (EO1385--0040047),
 detected with the Automatic Plate Measuring Machine.
In this case, the magnitude in the $R$ band is $17.56$, while in the
$B$ band it is $19.76$.

At $\sim10''$ from NGC4565--ULX3, we find the dust cloud NGC4565--D--064--016 (Howk \& Savage 1999). The cloud has
dimensions $210\times 400$~pc and a column density $N_H>2\times 10^{21}$~cm$^{-2}$. At the distance of $9.7$~Mpc, the angular
dimensions are $4.5'' \times 8.5''$, so that this association could be possible.

Inside the error box of NGC4565--ULX4, we find two objects:
one is the planetary nebula NGC4565--19, separated by $8.4''$ (Jacoby et al. 1996),
and the other is the globular cluster KAZF~$4565-7$, separated by $8.6''$ (Kissler--Patig et al. 1999).

Both ULX in NGC4639 are close to H~II regions (see Evans et al. 1996).
NGC4639--ULX1 is close to the region NGC4639--07 ($9.7''$ angular separation),
while NGC4639--ULX2 is near NGC4639--64 ($2.2''$) and NGC4639--81 ($8.2''$). The angular areas of
these H~II regions are $3.9$~arcsec$^2$, $1.4$~arcsec$^2$, $2.5$~arcsec$^2$, respectively.

For NGC4698--ULX1 we found radio counterparts with the VLA at 6~cm, with an approximate flux density
of $0.8$~mJy (Ho \& Ulvestad 2001). Also in the optical band there are counterparts detected
with the DSS (EO0041--0230928) and by the {\it Hubble Space Telescope}.
A more detailed study on this source is in preparation (Foschini et al., in preparation).

\section{Overall view and discussion}
We have analyzed the X--ray data from \emph{XMM--Newton} observations of 10 nearby Seyfert galaxies.
The host galaxies are located between $7.4$ and $21.3$~Mpc, with 7/10 between $16.8$ and $21.3$~Mpc.
Only one host galaxy (NGC4168) is elliptical (E2), while all the remaining are spirals of various types
(see Table~\ref{tab:host}).

We found ULX in 9 of the 10 galaxies. The contamination with background
sources is very low. Indeed, from \emph{XMM--Newton} observations of the Lockman Hole, Hasinger et al. (2001) found
about $100$ sources per square degree with flux higher than $10^{-14}$~erg cm$^{-2}$ s$^{-1}$ in the energy band
$0.5-2$~keV, and about $200$ sources per square degree in the energy band $2-10$~keV. Assuming the same
$\log N - \log S$ and considering the flux limit reached by our observations, we expect to find, in the worst case
(NGC4565), fewer than one (0.7) background source inside the $D_{25}$ ellipse. For all remaining host galaxies, the expected number of
background objects is significantly less than one (0.2 for most of the cases). Therefore, we expect that, in the worst case, the
overall sample contains fewer than 2 background objects.

The total number of ULX in the present catalog is 18. The mean value is 1.8 ULX per galaxy, with
the maximum value in NGC4565 with 7 ULX and the minimum in NGC4138 with no detection. By omitting NGC4565, we have a mean
value of about 1.2 ULX per galaxy. With respect to the {\it ROSAT}/HRI survey by Roberts \& Warwick (2000), we find that
\emph{XMM--Newton} allows a significative improvement in the number of 
ULX detections (see Table~\ref{tab:inc}). In Table~\ref{tab:inc} we list also 
the luminosities in $B$ and far-infrared bands of the host galaxies, the latter being a rough
indicator of the star formation activity. At a first look, no 
evident correlation appears, but the present sample is very small. Although it
appears that host galaxies with high $L_{\mathrm{B}}$ and $L_{\mathrm{FIR}}$ (NGC4501 and
NGC4565) have a higher number of ULX, we caution that this effect could be
spurious because our survey does not reach a uniform luminosity threshold
to detected ULX.  NGC4565, for example, reaches a flux limit of
$\sim$$1\times 10^{-14}$~erg cm$^{-2}$ s$^{-1}$, which is deeper than
the flux corresponding to the ULX luminosity limit of
$2\times 10^{38}$~erg s$^{-1}$ (cf. Sect.~2). For NGC4501, on the other hand,
the flux limit is too shallow to reach this luminosity limit.

\begin{table}[!ht]
\caption{Predicted and observed number of ULX. Columns: (1) Name of the host galaxy;
(2) total absolute $B$ magnitude, from Ho et al. (1997a); (3) luminosity [$10^{10}$~$L_{\sun}$~erg s$^{-1}$] calculated
from data in Col.~2; (4) expected number of ULX according to Roberts \& Warwick (2000), who found a relationship
between the number of ULX and $L_{\mathrm{B}}$; (5) number of ULX actually found in the present survey; (6)
luminosity in the far-infrared (FIR, $42.5-122.5$~$\mu$m) in units [$10^{42}$~erg s$^{-1}$], calculated from data of Ho et al. (1997a).}
\centering
\begin{tabular}{lccccc}
\hline
Galaxy & $M_{\mathrm{B}}$      & $L_{\mathrm{B}}$      & Expected   & Found & $L_{\mathrm{FIR}}$\\
(1)    & (2)          & (3)          & (4)        & (5)   & (6)\\
\hline
NGC1058 & $-18.25$      & 0.15        & 0.1        & 2    & 1.9\\
NGC3185 & $-18.99$      & 0.30        & 0.2        & 1    & 5.1\\
NGC3486 & $-18.58$      & 0.21        & 0.1        & 1    & 2.7\\
NGC3941 & $-20.13$      & 0.87        & 0.6        & 1    & 5.3$^{\mathrm{*}}$\\
NGC4138 & $-19.05$      & 0.32        & 0.2        & 0    & 2.0$^{\mathrm{*}}$\\
NGC4168 & $-19.07$      & 0.32        & 0.2        & 1    & 0.37\\
NGC4501 & $-21.27$      & 2.5         & 1.7        & 2    & 48\\
NGC4565 & $-20.83$      & 1.7         & 1.2        & 7    & 10\\
NGC4639 & $-19.28$      & 0.40        & 0.3        & 2    & 4.1\\
NGC4698 & $-19.98$      & 0.70        & 0.5        & 1    & 1.5\\
\hline
\end{tabular}
\label{tab:inc}
\begin{list}{}{}
\item[$^{\mathrm{*}}$] Since no FIR data were available in Ho et al. (1997a), we calculate the
FIR luminosity according to the relationship $\log L_{\mathrm{FIR}}/L_{\mathrm{B}}=-0.792$ (Pogge \& Eskridge 1993).
\end{list}
\end{table}

The luminosities observed are in the range $(2 - 74)\times 10^{38}$ erg s$^{-1}$, depending on the model considered.
If we make the simplistic assumptions that the accretion is uniform and 
spherical, that the bolometric luminosity approximately equals the X--ray 
luminosity, and that the Eddington ratio is $1$, these luminosities correspond
to compact objects with masses between $1.5$ and $57$ $M_{\sun}$. However, the 
X--ray luminosity is generally only $30-40\%$ of the bolometric luminosity of 
the accreting sources (e.g., Mizuno et al. 1999). In addition, if the
Eddington ratio is in the range of $0.1-0.01$, as suggested by observations of 
Galactic black hole candidates (e.g., Nowak 1995), the mass range would
shift toward $10^3-10^4$ $M_{\sun}$.  Unless the sources are very young,
such high masses are difficult to explain for off--centre sources, because 
dynamical friction would tend to drag the objects toward the centre in less 
than the Hubble time (cf. Binney \& Tremaine 1987).

Therefore, as proposed by several authors, alternative scenarios must be considered. For example,
Makishima et al. (2000) proposed a Kerr black hole scenario: in this case, the luminosity produced by a spinning black hole can be
up to 7 times larger than in a Schwarzschild black hole. On the other hand, King et al. (2001) suggested that the matter could accrete
anisotropically: an anisotropic factor of $0.1-0.01$ reduces the values of the mass to those typically observed in
X--ray binaries in the Milky Way. Something similar has been suggested by Begelman (2002): in this case,
the presence of inhomogeneities in radiation pressure--dominated accretion disks, as a consequence of
photon--bubble instability, would allow the radiation to escape.
Finally, K\"ording et al. (2002) and Georganopoulos et al. (2002) suggested the possibility of relativistic beaming
due to the presence of jets coupled to an accretion disk. Both are based on the microquasar model
by Mirabel \& Rodriguez (1999).

The statistics of the present observations do not allow us to discriminate clearly between the different models, but
we can infer some useful hints from the eight sources, which gave sufficient counts for a spectral fitting.
In 5/8 cases the best-fit model is obtained with a simple power law with $\Gamma \approx 1.9-2.3$ (for an example of spectra,
see Fig.~\ref{fig4}). One of these five sources (NGC1058--ULX1) presents an almost flat spectrum ($\Gamma = 1.1\pm 0.3$).
For the remaining sources (2/8), we obtained a best fit with the black body model with $kT\approx 0.5-0.9$~keV.

It is known that the emission expected from a black hole X--ray binary is variable:
in the hard state, the spectrum is typically a power law with $\Gamma \approx 1.3-1.9$,
while in the soft state the spectral index increases up to about 2.5 and a soft component appears in
the X--ray spectrum (e.g., Ebisawa et al. 1996). Therefore, our sources could be black hole X--ray binaries in a hard or soft state.
Terashima \& Wilson (2002) proposed the existence of two different populations of
ULX, one characterized by soft thermal and the other by non--thermal X--ray emission.
A possible key to distinguish between the available hypotheses can be to perform time variability studies,
but current statistics are too low for such a study.

It is interesting to note that the MCD model, which has often been successful in the past for
ULX (e.g., Colbert \& Mushotzky 1999; Makishima et al. 2000) is never the best fit in our data.
Even when we obtain a reasonable fit with MCD, a simple black body model is statistically
better. This may be due to the low photon counts of the present spectra.

The unsaturated Comptonization (CST) model does not give acceptable fits, except for ULX4 in NGC4565,
for which it represents the second best fit, after the power law.

Our understanding of the nature of ULX is limited by the fact that, to date, 
counterparts at other wavelengths are quite rare (Roberts et al. 2001; Pakull \& Mirioni 2002; Wang 2002). In our sample, we note that one source
(NGC4698--ULX1) is detected in the radio (6 cm). NGC4168--ULX1 and NGC4639--ULX2 both show a highly
probable optical counterpart. In the second case, it is identified as a H~II region, which is a type of counterpart
frequently associated with ULX (Pakull \& Mirioni 2002). It is worth noting the case of NGC4656--ULX3 could be
obscured by, rather than correlated with, a dust cloud.
Finally, the probable counterparts of NGC4168--ULX1 and NGC4698--ULX1 appear 
to be considerably red objects, with $B-R > 2$ mag.

\section{Final remarks}
We have found 18 ULX in a sample of 10 nearby Seyfert galaxies. This is the first step of a larger
survey comprising 28 Seyfert galaxies with distances smaller than 22 Mpc to be observed
with \emph{XMM--Newton}.

A more detailed analysis will be presented in later papers, when the complete sample
of \emph{XMM--Newton} observations will be available, together with follow--up 
observations at other
wavelengths. In the meantime, the present X--ray catalog provides a basis 
for future X--ray and optical studies.

\begin{acknowledgements}
This work is based on observations obtained with \emph{XMM--Newton}, an ESA science
mission with instruments and contributions directly funded by ESA Member States and the USA (NASA).
This research has made use of the NASA's Astrophysics Data System Abstract Service and of the NASA/IPAC Extragalactic Database (NED), which is operated by the Jet Propulsion Laboratory,
California Institute of Technology, under contract with the National Aeronautics and Space Administration.
We acknowledge the partial support of the Italian Space Agency (ASI) to this research.
\end{acknowledgements}

\end{document}